\documentclass[twocolumn, prx, superscriptaddress,notitlepage]{revtex4-2}
\usepackage{graphicx}% Include figure files
\usepackage{dcolumn}% Align Table columns on decimal point
\usepackage{bm}% bold math
\usepackage[usenames,dvipsnames]{color}
\usepackage[most]{tcolorbox}
\usepackage{multirow}
\usepackage{gensymb}
\usepackage[normalem]{ulem}
\usepackage{CJK}
\usepackage{comment}
\usepackage[colorlinks, linkcolor=blue,anchorcolor=blue,citecolor=blue,urlcolor=blue]{hyperref}
\usepackage{amssymb}
\usepackage{pifont}
\usepackage{physics}
\usepackage{natbib}
\usepackage{xcolor}
\usepackage[per-mode=symbol]{siunitx}
\usepackage{color,soul}

\begin{document}
\begin{CJK*}{UTF8}{}
\title{Cavity-enabled real-time observation of individual atomic collisions}

\author{Matthew L. Peters}
\thanks{These authors contributed equally.}
\affiliation{
   MIT-Harvard Center for Ultracold Atoms and Research Laboratory of Electronics, Massachusetts Institute of Technology, Cambridge, MA 02139, USA}
\affiliation{
   Department of Physics, Massachusetts Institute of Technology, Cambridge, MA 02139, USA}
   
\author{Guoqing Wang \CJKfamily{gbsn}(王国庆)} 
\thanks{These authors contributed equally.}
\affiliation{
   MIT-Harvard Center for Ultracold Atoms and Research Laboratory of Electronics, Massachusetts Institute of Technology, Cambridge, MA 02139, USA}
\affiliation{
   Department of Physics, Massachusetts Institute of Technology, Cambridge, MA 02139, USA}

\author{David C. Spierings}
\thanks{These authors contributed equally.}
\affiliation{
   MIT-Harvard Center for Ultracold Atoms and Research Laboratory of Electronics, Massachusetts Institute of Technology, Cambridge, MA 02139, USA}
\affiliation{
   Department of Physics, Massachusetts Institute of Technology, Cambridge, MA 02139, USA}

\author{Niv Drucker}
\affiliation{
   Quantum Machines, Tel Aviv-Yafo 6721407, Israel}

\author{Beili Hu}
\affiliation{
   MIT-Harvard Center for Ultracold Atoms and Research Laboratory of Electronics, Massachusetts Institute of Technology, Cambridge, MA 02139, USA}
\affiliation{
   Department of Physics, Massachusetts Institute of Technology, Cambridge, MA 02139, USA}

\author{Yu-Ting Chen}
\thanks{Current Address: Institute for Quantum Computing and Department of Physics and Astronomy, University of Waterloo, Waterloo, Ontario N2L 3G1, Canada}
\affiliation{
   MIT-Harvard Center for Ultracold Atoms and Research Laboratory of Electronics, Massachusetts Institute of Technology, Cambridge, MA 02139, USA}
\affiliation{
   Department of Physics, Massachusetts Institute of Technology, Cambridge, MA 02139, USA}
\affiliation{ Department of Physics, Harvard University, 17 Oxford Street, Cambridge, MA 02138, USA}

\author{Vladan Vuleti\'c}
\email[]{vuletic@mit.edu}
\affiliation{
   MIT-Harvard Center for Ultracold Atoms and Research Laboratory of Electronics, Massachusetts Institute of Technology, Cambridge, MA 02139, USA}
\affiliation{
   Department of Physics, Massachusetts Institute of Technology, Cambridge, MA 02139, USA}

\begin{abstract}
Using the strong dispersive coupling to a high-cooperativity cavity, we demonstrate fast and non-destructive number-resolved detection of atoms in optical tweezers. We observe individual atom-atom collisions, quantum state jumps, and atom loss events with a time resolution of $\SI{100}{\micro\s}$ through continuous measurement of cavity transmission. Using adaptive feedback control in combination with the non-destructive measurements, we further prepare a single atom with $92(2)\%$ probability.
\end{abstract}

\maketitle

\end{CJK*}

Arrays of individual neutral atoms represent a promising platform for quantum information processing due to their scalability, arbitrary connectivity, and long coherence times ~\cite{manetsch_tweezer_2024,bluvstein_quantum_2022,bluvstein_logical_2024}. These features are enabled in large part by the simple trapping and high-fidelity fluorescence imaging of individual atoms within tweezer traps. While the fluorescence imaging onto a camera has the advantage of parallelism, i.e.~many traps can be imaged simultaneously, the imaging time is limited by the effective photon collection efficiency of free-space optics, which is typically 1-2\% {\cite{manetsch_tweezer_2024}}. Atom detection via a cavity \cite{thompson_observation_1992,munstermann_observation_2000,pinkse_trapping_2000,mckeever_state-insensitive_2003,goldwin_fast_2011} offers the advantage that the signal-to-noise ratio can exceed that of an ideal $4\pi$ fluorescence detector, enabling much faster detection. Using macroscopic or nanofiber cavities, fast and high-fidelity state readout of single atoms has been demonstrated ~\cite{deist_mid-circuit_2022, CavityDetectionMoehring, nayak_real-time_2019,vetsch_optical_2010,goban_demonstration_2012,liu_realization_2023}. For several atoms, controlled super- and subradiant effects in the coupling of the atoms to the cavity have been recently observed~\cite{yan_superradiant_2023,zhang_cavity_2024,li_tunable_2024}.

In this Letter, we demonstrate fast, non-destructive, state-resolving measurements enabled by a high-cooperativity ring cavity ~\cite{YuTingCavity,zhang_cavity_2024,li_tunable_2024}. 
The strong atom-light coupling (cooperativity $\eta=21$) gives rise to large dispersive cavity shifts by individual atoms even far from atomic resonance, which is used to distinguish different numbers of atoms in a single dipole trap or small tweezer array. With a time resolution of \SI{100}{\micro\s}, we directly observe individual inelastic collisions between pairs of atoms in an optical tweezer, which represents the mechanism that underpins the loading of at most one atom into a tweezer~\cite{CollisionsAndersen, CollisionsBrowaeys, CollisionsGould}. We also demonstrate a protocol that uses cavity measurement and adaptive feedback to realize quasi-deterministic loading of a single atom with 92(2)\% fidelity.

\begin{figure}
    \includegraphics[width=0.49\textwidth]{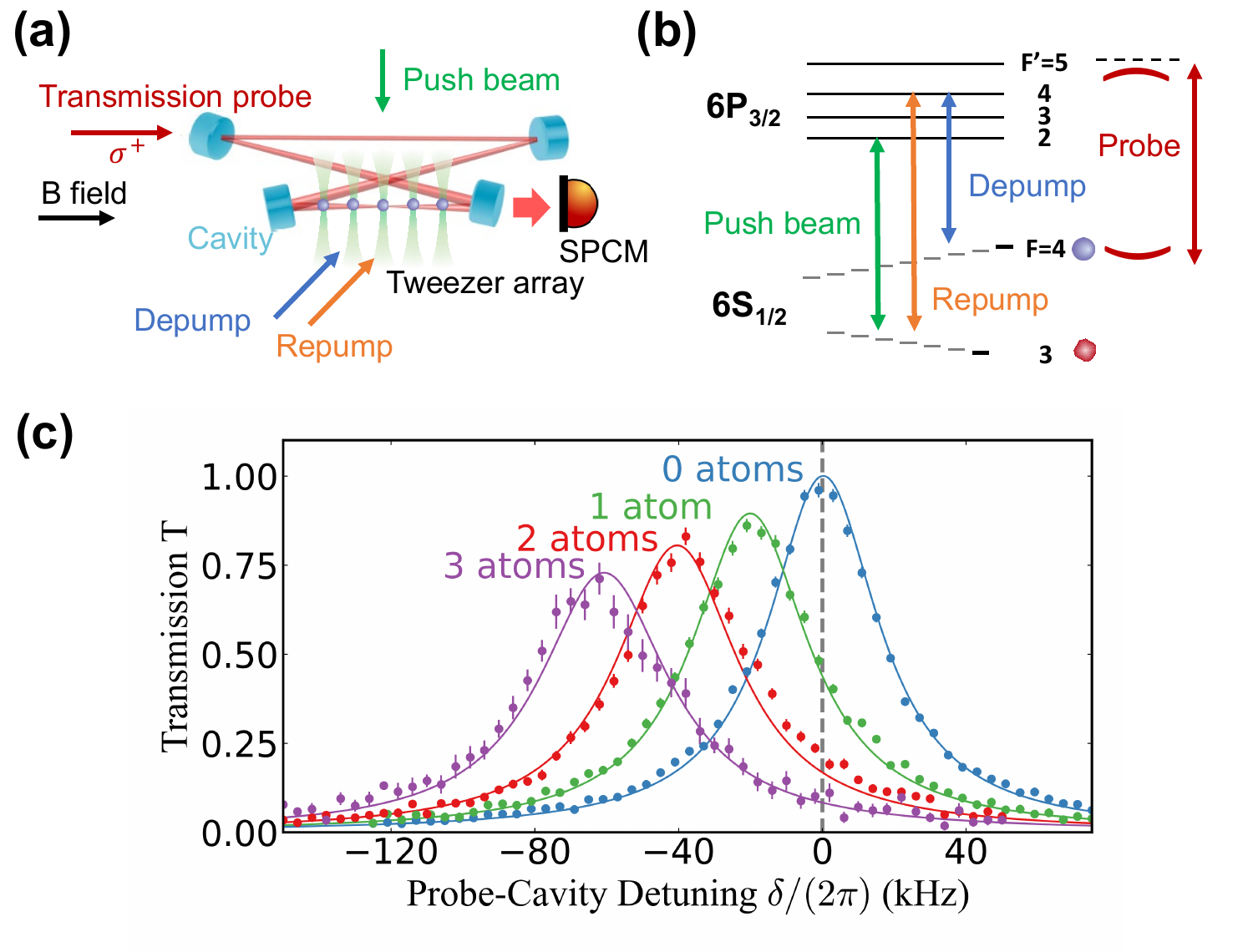} 
    \caption{{\bf Experimental configuration.} (a) Schematic of the experimental setup. Cesium atoms in optical tweezers are positioned within a bow-tie running-wave cavity. (b) Relevant level structure of Cs atoms. (c) Cavity transmission spectrum for 0 (blue), 1 (green), 2 (red), and 3 (purple) atoms trapped within the cavity. Here the cavity resonance is detuned by  $\Delta/(2\pi)=-50$~MHz from the $\ket{F=4}$ to $\ket{F'=5}$ component of the $D_2$ line. 
    }
    \label{fig:1}
\end{figure}

\begin{figure*}
    \includegraphics[width=\textwidth]{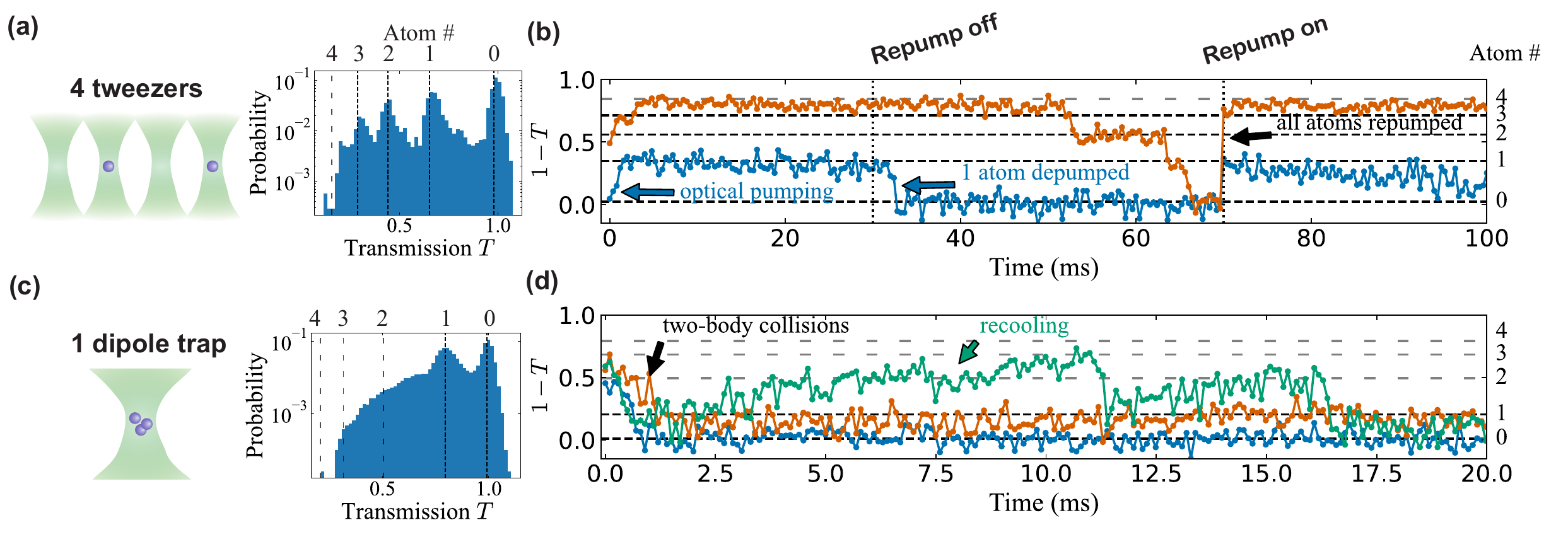}
    \caption{{\bf Real-time measurement of atom dynamics.} The cavity is detuned by $\Delta/(2\pi)=107$~MHz from atomic $\ket{F=4}\rightarrow \ket{F=5}$ resonance while the probe is resonant with the empty cavity ($\delta=0$). (a) Histogram of the probability for cavity transmission in 400-$\SI{}{\micro\s}$ bins for single atoms probabilistically loaded into four separate tweezers (waist $w=\SI{1.0}{\micro\m}$).
    Peaks in the distribution are associated with 0, 1, 2, 3 atoms (from right to left). (b) A measure of atom-cavity coupling $1-T$ for cavity transmission $T$, from the dataset in (a). Representative time traces showcase atom number resolution, as well as quantum jumps associated with optical pumping between the hyperfine ground states $F=3,4$. 
    (c,d) Measurements as in (a,b), but for several atoms initially loaded into a slightly bigger dipole trap (waist $w=\SI{1.4}{\micro\m}$) for 100-\SI{}{\micro\s} bins. Inelastic collisions cause heating or trap loss and decrease the coupling.
    }
    \label{fig:Traces}
\end{figure*}

Figure \ref{fig:1} illustrates the experimental setup. Cold Cs atoms are prepared in a magneto-optical trap (MOT) near the center of an in-vacuum bow-tie ring cavity, which has a roundtrip length of \SI{18.6}{\cm}, a waist of $w_c=\SI{7}{\micro\m}$, and a finesse of $\mathcal{F}=\SI{50000}{}$~\cite{YuTingCavity}. Atoms are loaded directly from the MOT into a one-dimensional array of optical tweezers generated by an acousto-optic deflector, and aligned along the axis of the cavity. The traps are created by 937-nm light, which is focused by an out-of-vacuum microscope objective (numerical aperture $NA=0.5$) to waists of $w=\SI{1.0}{\micro\m}$ with a trap depth $U/h=\SI{24}{\MHz}$ and radial trapping frequency $\omega_r/(2\pi)= \SI{85}{\kHz}$. The hyperfine state of the atoms is controlled by applying `repumping' and `depumping' light on the $\ket{6S_{1/2}, F=3} \rightarrow \ket{6P_{3/2}, F'=4}$ and $\ket{6S_{1/2}, F=4} \rightarrow \ket{6P_{3/2}, F'=4}$ hyperfine transitions, respectively.

We calibrate the strength of the atom-cavity coupling via the dispersive shift of the cavity resonance by individual atoms. Single atoms are prepared in the tweezer traps by a brief stage of polarization gradient cooling. This induces light-assisted collisions, leaving each tweezer with an occupation of zero or one atom~\cite{LightAssistedGrangier}.  The cavity transmission is then probed using $\sigma ^+$ polarized light, with a magnetic field of \SI{5.3\pm0.1}{G} applied along the cavity axis. 
Fig.~\ref{fig:1}(c) shows the measured cavity transmission for different numbers of individually trappped atoms, where the atom number is determined independently by imaging the atoms through the microscope objective onto a camera. For the data in Fig.~\ref{fig:1}(c), the cavity (frequency $\omega_c$) is detuned by $\Delta=\omega_c-\omega_a=- 2\pi \times 50$~MHz from the atomic $\ket{F=4} \rightarrow \ket{F'=5}$ resonance (frequency $\omega_a$), and the probe-cavity detuning $\delta=\omega_p -\omega_c$ is varied.

The single-atom cooperativity for coupling to the $TEM_{00}$ mode, $\eta=21.0(3)$, is extracted from a simultaneous fit of all the data to the transmission \cite{VladanClassical}
\begin{equation}
\label{eq:transmitted}
     T= 
     \frac{1}{\left(1+N \eta \frac{1}{1 + y^2} \right)^2+ \left(x-N \eta \frac{y}{1+y^2} \right)^2},
\end{equation}
where $N$ is the atom number in the cavity, $y=2(\omega_p-\omega_a)/\Gamma$ and $x=2(\omega_p-\omega_c)/\kappa$ are the normalized probe-atom detuning and probe-cavity detuning, with $\Gamma/(2\pi)=\SI{5.2}{\MHz}$ and $\kappa/(2\pi)=\SI{37}{\kHz}$ being the atomic and cavity linewidths, respectively.

%\textit{Real-time atom counting.---}
We operate in the dispersive regime of cavity probing ($N \eta/(1+y^2) < 1$), where the atomic absorption is small. For the cavity probe fixed on the empty-cavity resonance ($\delta=0$), we observe a significantly shorter trap lifetime during probing for $\Delta<0$ (typically 10~ms) than for $\Delta>0$ (typically 100~ms), which we attribute to cavity heating and cooling~\cite{CavityCoolingRitsch,vuletic_laser_2000,vuletic_three-dimensional_2001,CavityCoolingMurr}, respectively. Cavity cooling (heating) occurs if the cavity is blue (red) detuned relative to the probe beam. For $\delta=0$, this occurs when $\Delta>0$ ($\Delta<0$) due to the atom-induced dispersive shift (see also Fig.~\ref{fig:1}(c)). Photons scattered into the cavity then have a higher (lower) energy than the driving light, which results in a reduction (increase) of the atom's kinetic energy.
To observe the atoms while also cooling them, we operate at $\Delta/(2\pi)  = \SI{107}{\MHz}>0$, where the transmission is sufficiently atom number dependent to allow us to resolve 0, 1, 2, and 3 atoms within a typical measurement time of \SI{100}{\micro\s}. 

Figure~\ref{fig:Traces}(a) shows the histogram of the measured cavity transmission $T$, which displays pronounced peaks (indicated by the dashed lines) corresponding to different atom numbers inside the cavity. Those peaks provide a calibration of atom number when we observe individual time traces displaying real-time atomic dynamics, where the short-dashed lines are extracted from the peak positions and the long-dashed lines are extrapolated from the measured cooperativity. (For atoms trapped in an intracavity lattice, similar continuous-time signals were first observed in Ref.~\cite{McKeeverCavityProbing}.) Fig.~\ref{fig:Traces}(b) shows characteristic time traces when we probe the cavity coupling $1-T$ after probabilistic loading of four separate tweezers, each containing at most one atom.
The step-like time traces in Fig.~\ref{fig:Traces}(b), display quantum jumps associated with the optical pumping of individual atoms between hyperfine ground states, since only atoms in the manifold $\ket{F=4}$ are strongly coupled to the cavity mode. The total atom number is measured by turning on the repumping light that transfers all atoms to the $\ket{F=4}$ manifold.
At the beginning of each trace, we observe a fast decrease in transmission, with a measured time constant of \SI{1.1\pm0.2}{\ms}, corresponding to the optical pumping of atoms to the $\ket{F=4,\ m_F=4}$ magnetic sublevel, for which the coupling to the circularly polarized cavity mode is maximized.

Having calibrated and verified the atom number dependent transmission, we then proceed to probe the dynamics of several atoms in a single trap (Fig.~\ref{fig:Traces}(c)). By removing the polarization gradient cooling from the loading procedure and slightly enlarging the trap waist size to $w=\SI{1.4}{\micro\m}$ (trap depth {$U/h=\SI{22}{\MHz}$}, radial trapping frequency $\omega_r/(2\pi)=\SI{58}{\kHz}$, and axial frequency $\omega_{ax}/(2\pi)=\SI{12}{\kHz}$), we load multiple atoms into a single dipole trap. In the larger trap the atomic density and rate of light-induced two-body collisions \cite{LightAssistedGrangier,RegalCollisions} are sufficiently reduced to become observable by the cavity measurement. Figure {\ref{fig:Traces}(d)} shows representative multi-atom time traces that display random large and abrupt transmission changes consistent with collisional loss and collisional heating induced by binary atomic collisions within the single trap. We observe that when the trap initially contains two atoms, a collision event often leaves the trap empty, while a collision in a sample of three atoms often results in a single, stably trapped atom. We conclude that we are observing the real-time collisional dynamics upon which single-atom tweezer loading relies~\cite{LightAssistedGrangier,RegalCollisions}.

We also observe instances of collisions that result in temporarily reduced coupling, but not in the ejection of atoms from the dipole trap. For example, the green trace in Fig.~\ref{fig:Traces}(d) shows a sharp decrease of coupling $1-T$ at $t_1$=1~ms, followed by a gradual increase of the atom-cavity coupling back to the value prior to the abrupt change. We interpret this behavior as representing a collision that imparts a significant amount of kinetic energy to the atoms, though insufficient to eject them from the trap. Atoms with a high kinetic energy following an inelastic collision are displaced further from the center of the cavity mode (waist $w_c=\SI{7}{\micro\m}$), and more weakly coupled. As cavity cooling reduces the atomic temperature, the coupling slowly increases, as observed in the green trace for times between 2 and 11~ms, before two more inelastic collision events, one at $t_2=\SI{11.5}{ms}$ and one at $t_3=\SI{16.5}{ms}$, expel two atoms, and leave just one atom remaining in the trap. (We observe that single atoms remain trapped for long times, with no significant changes in coupling.) As a consequence of the changes of transmission due to collisions and recooling, the histogram for several atoms in a single trap, Fig. \ref{fig:Traces}(c), only displays two peaks, corresponding to the stable traces for zero and one atoms.
We note that optical pumping (see Fig. \ref{fig:Traces}(b)), which can also lead to increased cavity coupling, occurs on a faster ($\sim 1$~ms) time scale.

\begin{figure}[h]
    \includegraphics[width=0.48\textwidth]{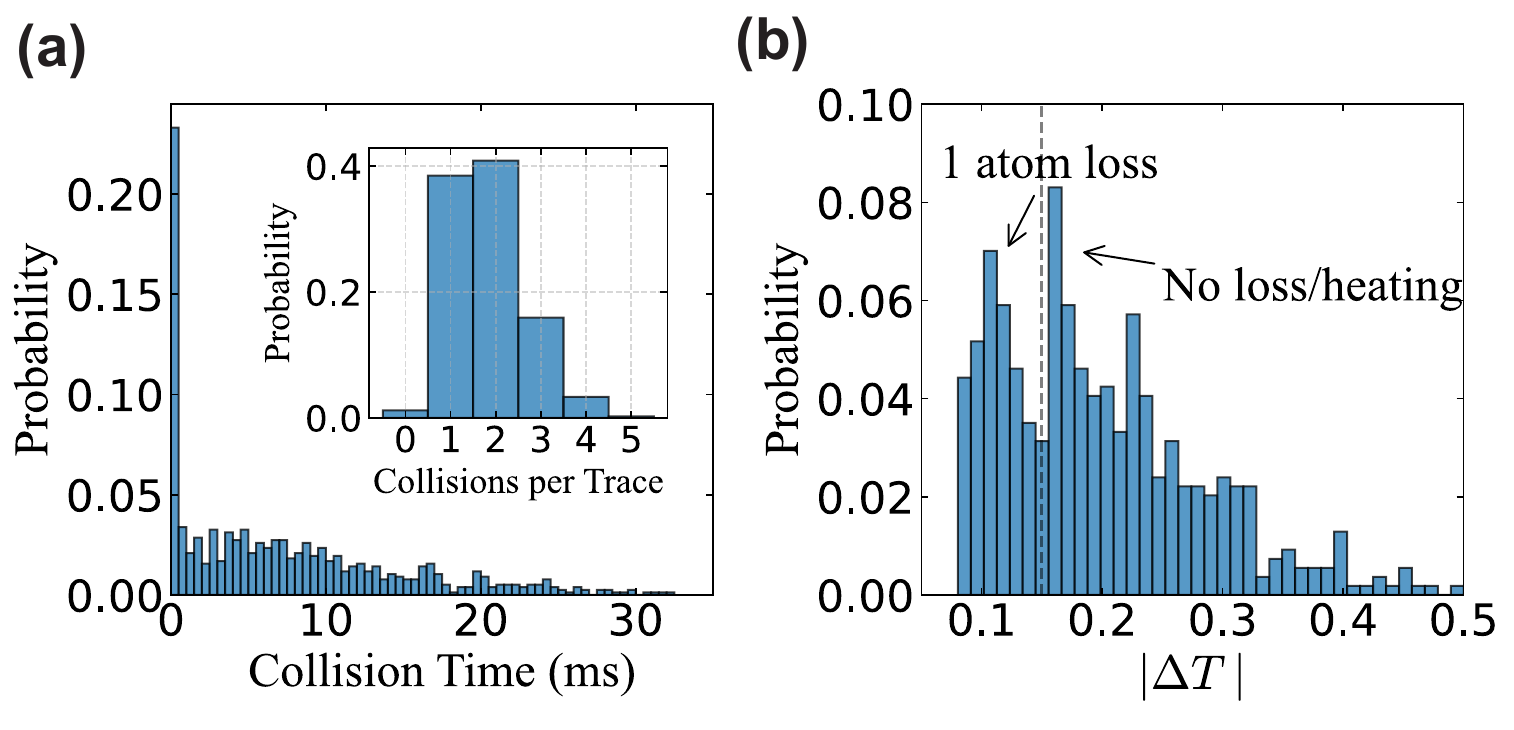}
    \caption{{\bf Analysis of atomic collisions.}
    (a) Probability distribution of the collision times from time traces as in Fig.~\ref{fig:Traces}(d). Several atoms are loaded at $t=0$ into a single dipole trap. We identify atomic collisions as abrupt increases in transmission that correspond to an apparent loss of more than one atom. The inset shows the number of observed atomic collisions before a steady state with zero or one atoms is reached. (b) Histogram of the slow increase in cavity coupling $1-T$ that follows 50\% of the detected atomic collisions. We observe a peak consistent with retention of one atom after collisions ($|\Delta T| \sim 0.12$), as well as greater increases in coupling ($|\Delta T| > 0.15$) indicating no loss of atoms. Changes $|\Delta T| \leq 7$\% are consistent with shot noise and not considered.
    }
    \label{fig:collisions}
\end{figure}

We further analyze the collisional dynamics and summarize our findings in Fig.~\ref{fig:collisions}. Our data reveal that 20\% of the observed collisions occur within the first \SI{100}{\micro\s} of cavity probing, while the remaining collisions occur over a much longer timescale of $\sim 9$~ms (see Fig.~\ref{fig:collisions}a).
We attribute the fast initial collision rate as being due to a larger atomic density immediately after trap loading. The longer timescale probably arises from multiple axial oscillations along the tweezer, where two atoms with a collision cross section $\sigma \approx (\lambda/(2\pi))^2$ with $\lambda=852$~nm need to find each other. The inset in Fig.~\ref{fig:collisions}(a) displays the number of collisions observed per experimental cycle before we end up with zero or one atoms. For our conditions, on average 1.8 collisions are needed to reduce the atom number to zero or one.

Our data can also be used to characterize the amount of heating in an individual inelastic collision. As described above, we observe time traces that correspond to two types of collisions: traces with a sudden decrease in coupling to a new constant level can be associated with the colliding atoms leaving the trap, while traces with a sudden decrease followed by a slow increase in cavity coupling, either to the original level or to a lower level, can be attributed to one or both atoms remaining in the trap and being recooled to the cavity mode center. Fig.~\ref{fig:collisions}(b) quantifies this increase in cavity coupling that often follows a collision: the distribution of coupling increases shows distinct peaks. For instance, when only a small amount of coupling ($\sim 10$\%) is regained after a collision, we interpret this to indicate that one of the colliding atoms has left the trap, whereas a larger coupling gain likely indicates the retention and recooling of both atoms.
We find that about half of the observed collisions are followed by an increase in cavity coupling, suggesting the retention of one or both atoms, while the other half of the observed collisions correspond to the loss of both atoms. The sum of the probabilities below the vertical dashed line in Fig.~\ref{fig:collisions}(b) provides an upper-bound of $\sim15\%$ probability to lose one atom after a collision. Collisions that result in the loss of only one atom are known to result in an average loading probability for one atom of over 50\% \cite{brown_gray-molasses_2019,grunzweig_near-deterministic_2010,fung_efficient_2015,lester_rapid_2015}. In our experiment, collisional blockade loads a single atom 55\% of the time, consistent with only a slight bias from the single atom retention.

Finally, we implement a cavity-measurement-based adaptive feedback protocol for quasi-deterministic loading of a single atom starting with a small ensemble. As before, we load several atoms into either several traps or a single trap, and optically pump them to the $F=3$ hyperfine manifold. We then apply a weak repumping pulse to the $F=4$ manifold, so that on average less than one atom is transferred. Each repumping stage is followed by a 1-ms long measurement of atom number in the $F=4$ manifold via the cavity. If a single atom is successful repumped, the remaining atoms in $F=3$ are pushed out of the trap by an intense laser beam tuned to the $\ket{F=3} \rightarrow \ket{F'=2}$ transition. If no atoms are found in $F=4$, repumping is attempted again, whereas if more than one atom is observed, all atoms are depumped to $F=3$, and the procedure is restarted (see Supplemental Materials~\cite{SOM} for details of the adaptive algorithm).
%Figure \ref{fig:4}(b) shows an example sequence following the adaptive logic in which a single atom was successfully loaded after the second repumping attempt.

\begin{figure}
    \includegraphics[width=0.495\textwidth]{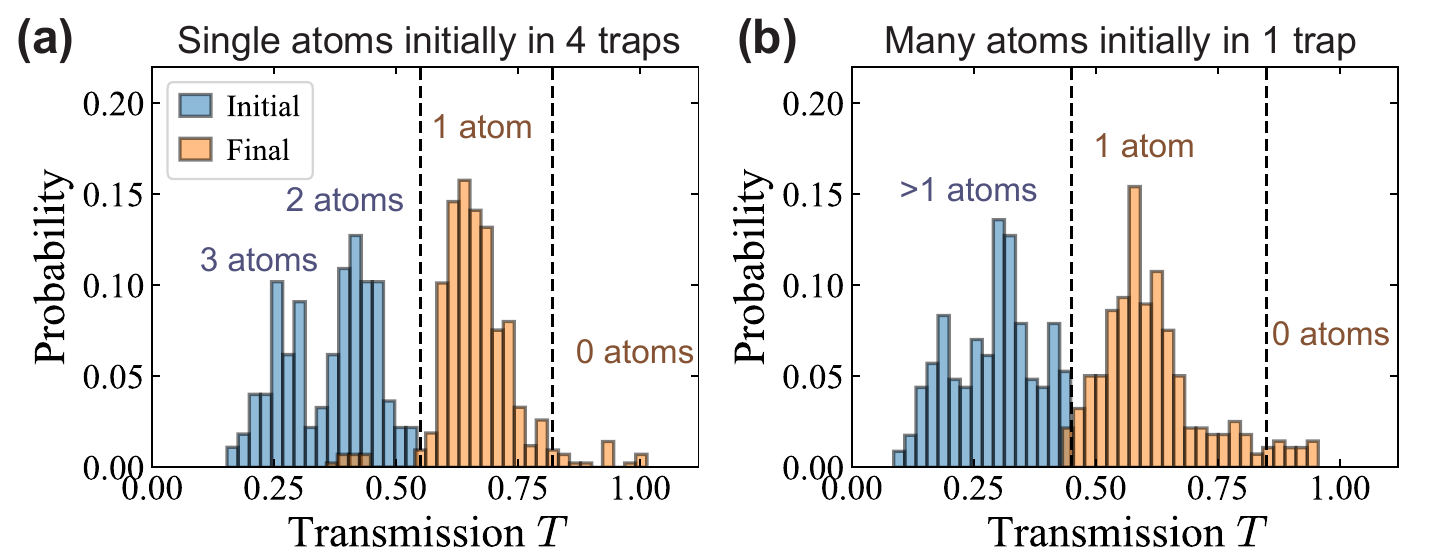} 
    \caption{{\bf Adaptive loading of a single atom.}
    The probability to successfully load a single atom in the adaptive protocol (see text) is demonstrated for initial loading of at least two atoms in (a) four tweezer traps (275 trials, cavity-atom detuning $\Delta/(2\pi)=-\SI{73}{\MHz}$) and (b) one dipole trap (228 trials, $\Delta/(2\pi)=-\SI{58}{\MHz}$). The initial transmission histogram is shown in blue, while the orange histogram shows the final transmission. The two dashed lines set the threshold for the single-atom transmission. Final-transmission data falling outside the threshold correspond to failures due to atom loss, ejection errors, or reaching the limit of 50 attempts. 
    }
    \label{fig:4}
\end{figure}

Figures \ref{fig:4}(a) and \ref{fig:4}(b) show the performance of the adaptive procedure when preparing a single atom starting from several atoms in four tweezers and in a single dipole trap, respectively. The multi-trap and single-trap datasets use atomic detuning of $\Delta/(2\pi)=-78$~MHz and $-58$~MHz, respectively, in order to more clearly distinguish cavity transmission for one atom from that for other atom numbers. To avoid trivial procedures where we start and end with one atom or no atom at all, we postselect the datasets on having an initial atom number greater than or equal to 2, as measured by the cavity transmission at the end of the initial optical pumping. The success probability of the adaptive procedure is 92(2)\%, with an average time to success of about \SI{15}{\ms}, both when loading from four tweezers and from a single trap. The similar performance of the adaptive protocol in these two cases demonstrates that the pulsed interrogation protocol can minimize collisional loss.
 
The success of the adaptive protocol is primarily limited by the incomplete selective ejection of atoms in the $F=3$ manifold, which is only successful 80\% and 63\% of the time for the multiple traps and single trap, respectively. As many ejection failures leave multiple atoms in $F=4$, the adaptive protocol can simply try again, allowing the overall success probability to still remain high. Failed ejection does, however, increase the average time to success and thereby leaves more time for atom loss events, which are unrecoverable. We estimate that with improved ejection, both datasets in Fig.~\ref{fig:4} could have attained success probabilities of 99\%. Ejection of population in one hyperfine manifold via atom ejection on a stretched-state cycling transition has been demonstrated with efficiencies of $>98 \% $~\cite{BlowoutBirkl, BlowoutBrowaeys, BlowoutGrainger, BlowoutMeschede}.

The techniques demonstrated here could be expanded in several directions. 
Adaptive feedback can be combined with machine learning algorithms to enhance speed and efficiency, as recently used in the laser cooling to Bose-Einstein condensation~\cite{xu_bose-einstein_2024} and entanglement-enhanced sensing with spin squeezed states~\cite{duan_continuous_2024}. This could enable fast, parallelized, and near-deterministic loading of single atoms in cavity arrays without rearrangement~\cite{simon_cavity_array}. Dynamic adjustment of the atomic detuning in our protocol could allow for the clear resolution of two or three atoms, rather than one, allowing for studies of molecular formation \cite{CavityMoleculeDenschlag} or well-controlled light-assisted collisions~\cite{RegalCollisions}. Adaptive feedback also holds promise for mid-circuit operations for quantum computation~\cite{singh_mid-circuit_2023-1,deist_mid-circuit_2022,CavityDetectionMoehring,graham_midcircuit_2023,ma_high-fidelity_2023-1,norcia_midcircuit_2023,lis_midcircuit_2023}. Future implementations of adaptive protocols could also further improve performance by utilizing cavities optimized for readout and preparation speed.

In summary, we have realized nondestructive atom counting in real time through measurements of cavity transmission using a high-cooperativity bow-tie cavity. This measurement enables continuous, time-resolved probing of
two-body collisional dynamics of atoms within the same trap. Our work opens up many new opportunities from the fundamental atomic physics of cold atom collisions and chemical reactions to advancements of neutral-atom quantum information processors.

\acknowledgements
This work is supported by the U.S. Department of Energy, Office of Science, National Quantum Information Science Research Centers, Quantum Systems Accelerator. Additional support is acknowledged from the NSF Frontier Center for Ultracold Atoms (grant number PHY- 2317134), the NSF QLCI Q-SEnSE (grant number QLCI-2016244), 
and the DARPA ONISQ programme (grant number 134371-5113608).
We acknowledge Josiah Sinclair, Edita Bitikji, Gefen Baranes, and Michelle Chong for discussions of light-assisted collisions observed via cavities. The authors thank Alyssa Rudelis, Neng-Chun (Allen) Chiu, Michel Szurek, and Simone Colombe for initial work in constructing the experimental apparatus. We additionally thank Berk Kovos and Christian Hahn for their help in discussing integrating Quantum Machines OPX+ devices into our setup.

\bibliography{main_text}

\onecolumngrid
\clearpage
\begin{center}
\textbf{\large Supplemental Material: Cavity-enabled real-time observation of individual atomic collisions}
\end{center}
\setcounter{section}{0}
\setcounter{equation}{0}
\setcounter{figure}{0}
\setcounter{table}{0}
\setcounter{page}{1}
\makeatletter
\renewcommand{\theequation}{S\arabic{equation}}
\renewcommand{\thesection}{S\arabic{section}}
\renewcommand{\thefigure}{S\arabic{figure}}

\section{Pushing Atoms out of Traps}

In our adaptive measurements, we utilize a push beam directed along the longitudinal axis of our optical dipole trap and tuned to the $F=3 \rightarrow F'=2$ transition to selectively heat atoms in the $F=3$ mainfold out of the trap without influencing atoms in the $F=4$ manifold. Several considerations are essential for optimizing this process. First, atoms in the $F=3$ manifold can be inadvertently pumped to the $F=4$ manifold before they are heated out of the trap. To mitigate this, we minimize depumping to $F=4$ by tuning the laser frequency to $F=3 \rightarrow F'=2$ resonance and maintaining the laser intensity below the saturation threshold. Additionally, atoms may decay into a state in the $F=3$ manifold that are dark to the push beam. To counteract this, we make the push beam $\sigma^+$ polarized along the direction of propagation and direct it orthogonally to the quantization axis, ensuring all polarization components are present. We also apply a magnetic field strong enough to induce a magnetic splitting of 5.3(1) MHz between adjacent $m_F$ states in the $F=3$ manifold to couple the atom to the bright state when it does decay to the dark state. However, the magnetic field broadens the transition and there is a tradeoff between quickly rotating the atom out of the dark state and minimizing depumping, which we believe may be the limit on our ejection fidelity. We experimentally optimize the push process by tuning the laser frequency, laser power, and magnetic field strength to maximize the probability of heating atoms out of the trap. 
We calculate these efficiencies by dividing the number of unsuccessful push out attempts by the total number of attempts in our adaptive sequence. 

The push out scheme in our work utilizes the $F=3 \rightarrow F'=2$ transition in contrast to the typical $F=4 \rightarrow F'=5$ cycling transition used in most other push schemes due to our cavity probe  being on the $F=4 \rightarrow F'=5$ transition. The optical dipole traps in our experiment are magic-wavelength traps that shift the excited $6 P _{3/2}$ state by the same amount as the ground $6 S _{1/2}$ state. This eliminates any contributions to heating from dipole force fluctuations, which are often detrimental to experiments, but in this context, would be beneficial. Our push out efficiency could be improved if there were dipole force fluctuations for an anti-trapped excited state. Alternatively,  atoms in the $F=3$ manifold could be selectively excited to a Rydberg state, where they will be anti-trapped. Implementation of either of these would enhance the probability of a successful ejection and therefore our single-atom loading probability.

\section{Adaptive Feedback Details}
\begin{figure}
    \includegraphics[width=0.8\textwidth]{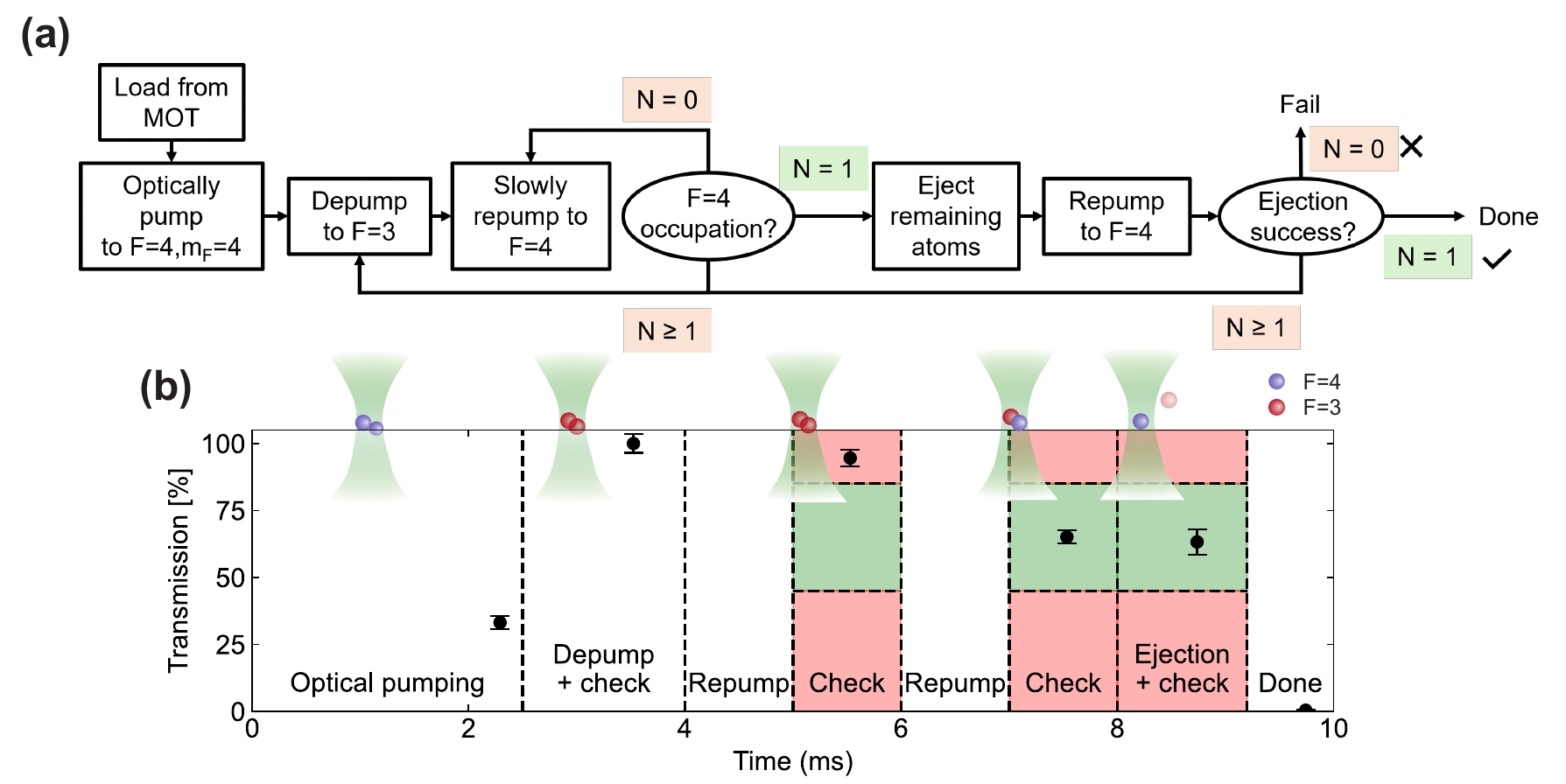} 
    \caption{{\bf Adaptive loading sequence and an example trace.}
    (a) Decision tree illustrating the logic that determines the pulse sequence in the adaptive feedback loading protocol. Following initial loading of atoms in the $\ket{F=4, m_F = 4}$ state, repetitive depumping and weak repumping between the $F=3$ and $F=4$ hyperfine manifolds occurs until a single atom is detected by the dispersive cavity probe. Afterwards, ejection of atoms remaining in $F=3$ is attempted. 
    (b) A representative successful trial, in which a single atom is loaded after one depumping stage and two repumping stages.
    }
    \label{fig:3_supp}
\end{figure}

The protocol for adaptive loading is shown in Fig.~\ref{fig:3_supp}(a), and an experimental example of a successful trail is shown in Fig.~\ref{fig:3_supp}(b). We utilize a Quantum Machines OPX+ to implement the adaptive logic described in this work.  Each adaptive action (listed in the bullets below) is followed by a measurement of atom number in the $F=4$ manifold with a duration of \SI{1}{\milli\second}. For each iteration in the adaptive protocol, the measured atomic number informs the next action based on a probability distribution function that correlates with the number of atoms in the cavity. Immediately following \textbf{Optical Pumping}, \textbf{Depumping}, and \textbf{Weak Repumping}, the atom check uses no repumper, only $F=4\rightarrow F'=5$ cavity probe light is used, whereas for \textbf{Push Out}, a strong repumper is used during the atom check to ensure that there is only a single atom remaining in both the $F=3$ and $F=4$ manifolds. The adaptive agent chooses the next action by measuring the mean probe transmission normalized to the total power for every sequence to eliminate long-term drifts. If it identifies more than one atom, \textbf{Depumping} will be chosen as the next action and the sequence begins again.

The adaptive protocol incorporates the following possible choices the agent has after each atom number measurement:

\begin{itemize}
    \item \textbf{Optical Pumping:} $\sigma ^+$ cavity probing light tuned to the $F=4 \rightarrow F'=5$ to optically pump atoms, and a duration of \SI{5}{\milli\second} is used to initialize the atoms in the stretched state.
    \item \textbf{Depumping:} Light tuned to the $F=4 \rightarrow F'=3$ is used to transfer all atoms out of the hyperfine manifold coupled to the cavity. A duration of \SI{500}{\micro\second} is used. During this depumping process, atoms scatter few photons and remain near higher $m_F$ levels, so that when repumped, they stay close to the stretched state.
    \item \textbf{Weak Repumping:} Light tuned to the $F=3 \rightarrow F'=3$ is used to slowly bring atoms, nominally one at a time, into the $F=4$ hyperfine manifold coupled to the cavity. The duration of weak repumping is \SI{1}{\milli\second} and the pumping rate into the $F=4$ manifold is \SI{2.63}{\milli\second}$^{-1}$.
    \item \textbf{Push Out:} Light tuned to the $F=3 \rightarrow F'=2$ is used to selectively remove atoms in the $F=3$ manifold via recoil heating. The duration of the push out is \SI{200}{\micro\second}.
\end{itemize}

After \textbf{Weak Repumping}, if the mean of the transmission in the atom check following the action falls within the window for a single atom, the agent will then attempt to \textbf{Push Out} the remaining $F=3$ atoms. If the mean normalized transmission is above (below) this window and the previous step is \textbf{Weak Repumping}, it will enact \textbf{Weak Repumping} (\textbf{Depumping}) as the next action. After attempting to \textbf{Push Out} the $F=3$ atoms following a successful \textbf{Weak Repumping}, the agent will end the sequence if the next atom check, which for this scenario includes a strong repumper + cavity probe, identifies a single atom or no atom (the second case indicating all atoms were lost). 
 
 This approach ensures that the adaptive agent can dynamically respond to the current state of the system and optimize the sequence of actions to achieve high single-atom loading probabilities. The adaptive agent executes up to 25 iterations for single atoms in multiple traps and up to 50 iterations for multiple atoms in a single trap. The average time taken per cycle was 15.7 ms for the single atoms in multiple traps and 14.9 ms for ensembles, of which 5 ms is due to optically pumping the atoms. To reduce this time, the detection pulses in our adaptive sequence could likely be decreased in duration to tens of microseconds without sacrificing fidelity.

The total number of shots that met our postselection criteria of having at least 2 atoms was 275/386 for the many-traps case and 228/801 for the single-trap case. We note that due to technical constraints on our cavity and vacuum chamber, the number of cesium atoms loaded in our MOT is small and there are a non-negligible number of 0 and 1 atom loading events. 
As noted in the main text, the vast majority of single-atom loading failures are due to failures of our push beam. In fact, for both datasets, there are only 3 traces each that are not due to ejection failures. For the single traps, all three are because the adaptive algorithm terminated before reaching a single atom after 25 iterations. For the ensemble case, two are due to atom loss during optical pumping and one is due to loss of all atoms during the adaptive sequence.

\end{document}